\documentclass[prd,preprintnumbers,twocolumn,amsmath,amssymb,nofootinbib,showpacs]{revtex4}
\def\scn#1#2{\section{#1}\lb{#2}}
\def\sscn#1#2{\subsection{#1}\lb{#2}}
\usepackage{epsfig,amssymb,amsfonts,verbatim}
\def\bfl{\begin{flushleft}}
\def\efl{\end{flushleft}}
\def\bfr{\begin{flushright}}
\def\efr{\end{flushright}}
\def\bc{\begin{center}}
\def\ec{\end{center}}
\def\be{\begin{equation}}
\def\ee{\end{equation}}
\def\ba{\begin{eqnarray}}
\def\ea{\end{eqnarray}}
\def\baa#1{\begin{array}{#1}}
\def\eaa{\end{array}}
\def\bw{\begin{widetext}}
\def\ew{\end{widetext}}

\def\lb#1{\label{#1}}

\def\schrod{Schr\"odinger  }

\def\grsim{~^{>}_{\sim}}

\def\Sign#1{\, \text{sign}\left(#1\right) }

\newtheorem{Def}{Def.}[section]

\newtheorem{Prp}[Def]{Proposition}

\def\cmo{\mu}

\begin{document}

\preprint{Grav. Cosmol. 16 (2010) 288-297  [arXiv: 0906.4282]}

\title{
Logarithmic nonlinearity in 
theories of quantum gravity:\\
Origin of time and observational consequences
}

\author{Konstantin G. Zloshchastiev}

\affiliation{National Institute for Theoretical Physics (NITheP),
%Stellenbosch Institute for Advanced Study,
%Stellenbosch, South Africa}
%\affiliation{and Institute of Theoretical Physics, University of Stellenbosch, Stellenbosch 7600, South Africa}
%\affiliation{
Department of Physics and Centre for Theoretical Physics,
University of the Witwatersrand,
Wits 2050, Johannesburg, South Africa}

%\affiliation{Department of Physics, National University of Singapore,
%Singapore 117542 %, Republic of Singapore
%}

%\affiliation{Department of Theoretical Physics, Dnepropetrovsk National University, Dnepropetrovsk 49050, Ukraine}

%\date{~ ~~~~~~~~~~~~~~~~~~~~~~}
%\date{Submitted on 10 Mar 2009}
%\date{~Received: 26 May 2000 [PRL], 1 June 2000 [LANL] ~}
%\date{Received \today}

%\scriptsize%\footnotesize

\begin{abstract}
Starting from a generic generally covariant classical theory we introduce the logarithmic correction to the quantum wave equation. We demonstrate the emergence of the evolution time from the group of automorphisms of the von Neumann algebra governed by this non-linear correction. It turns out that such time parametrization is essentially energy-dependent and becomes global only asymptotically - when the energies get very small comparing to the effective quantum gravity scale. 
Similar thing happens to the Lorentz invariance - in the resulting theory it becomes an asymptotic low-energy phenomenon.
We show how the logarithmic non-linearity deforms the vacuum wave dispersion relations and explains certain features of the astrophysical data coming from recent  observations of high-energy cosmic rays. In general, the estimates imply that {\it ceteris paribus} the particles with higher energy propagate slower than those with lower one, therefore, for a high-energy particle the mean free path, lifetime in a high-energy state and, therefore, travel distance from the source can be significantly larger than one would expect from the conventional theory.
Apart from this, 
we discuss also the possibility and conditions
of the transluminal phenomena in the physical vacuum such as the 
Cherenkov-type shock waves.

%~\\ \textit{Key words}: Quantum gravity - elementary particles - relativity - gamma rays: bursts
\end{abstract}

%\keywords{ddd}
\maketitle

%\narrowtext

%\normalsize
%\newpage

\scn{Introduction}{sec-i}
In the conventional quantum mechanics the linearity of the wave equation
is something
which is implicitly presupposed, yet the possibility of the non-linear generalization
has not been ruled out by experiment \cite{Weinberg:1989us}.
From the theoretical point of view,
there exist arguments that a nonlinearity in general can lead to the violation
of locality via the Einstein-Podolsky-Rosen (EPR) apparatus \cite{Gisin:1989sx}.
However, it has been also pointed out that instead of this nonlinearities can lead
to communications between branches of the wave function, they can be large in a fundamental
theory yet be unobservably small when measured experimentally \cite{Polchinski:1990py}.
Afterwards it was also shown that the locality is not violated for product states 
for a large class of
nonlinear generalizations, including the one which is being discussed here,
see Ref. \cite{Czachor:1997pw}, and references therein, and the debates
continue.
In any case,
the linearity requirement becomes a rather strict
and unnecessary assumption
if one expects quantum mechanics to be valid on a wide range of scales: for instance, the
modern theory of quantum gravity is believed to be essentially non-linear -
because the propagating particle will cause the quantum fluctuations in gravitational
medium which will react back.

Some non-linear extensions of QM have been already proposed -
for instance, in Ref. \cite{Doebner:1996} authors
studied a family of the non-linear wave equations for non-relativistic QM associated
with unitary group of certain non-linear gauge transformations of third kind - those
which leave the positional probability density invariant.
Another  approach to including non-linearity into QM is based on generalizing
the quantum phase space to the class of K\"ahler manifolds which admit certain Hamiltonian flow \cite{Ashtekar:1997ud}.

On the other hand, those who want to add non-linear terms into the wave equation
inevitably arrive at the problem of choice: it seems that there exists a huge variety of the
non-linear corrections which can be added without undermining the pillars of
the conventional QM - the concepts of
the physical state, probability, observables and measurement \cite{Singh:2007dj}.
Thus, the necessity of non-linear corrections encounters
the practicality issue:
what are the before unsolved problems of
the conventional QM which can be cleared up by introducing non-linearity?

In present paper we do not   {\it ab initio} postulate the
linearity of quantum wave equation at all energy scales.
Instead we consider the including of one particular nonlinear term, 
the logarithmic one,
into the quantum wave equation while preserving the generally
covariant structure for the classical theory.
We know that, among other things,
it would not induce
correlations for non-correlated systems 
in the flat-spacetime limit
\cite{Czachor:1997pw,BialynickiBirula:1976zp}.
We show that this nonlinearity naturally transforms into the evolution time derivative
which
becomes global and energy-independent only in the low-energy limit - when the energies are small comparing to the effective quantum gravity scale.
Then we discuss the vast observational implications of our model.

\scn{Non-linearity and additivity}{sec-add}
The formal structure of a generally covariant quantum theory is as follows.
Let $\Gamma_{\text{ex}}$ be the space of solutions of the generally covariant
equations of motion endowed with a degenerate symplectic structure defined by these
equations.
The degenerate directions of this symplectic structure integrate in orbits and the solutions
which belong to the same orbit must be physically identified.
The orbits form a symplectic space $\Gamma$ - a fully covariant object which becomes the physical phase space of the theory.
%This space is a fully covariant object and is also isomorphic to the reduced ADM phase space.
%and thus the latter is a fully covariant object independent from the (unphysical) ADM hypersurface initially used for its construction.
The set $A=C^\infty (\Gamma)$ of the real smooth functions on $\Gamma$, called the physical
observables, form an Abelian multiplicative algebra.
These observables are regarded as classical limits of the non-commuting quantum observables
whose ensemble forms the non-Abelian $C^*$-algebra $\cal{A}$.
Since $A$ is a non-Abelian algebra under the Poisson bracket operation one can assume that
$\cal{A}$ is a deformation of a subalgebra of the classical Poisson algebra.
Note that in $\cal{A}$ in general there is no defined Hamiltonian evolution or
representation of the Poincar\`e group, therefore, the time evolution is only determined by the dependence
of the observables on clock times.
% \cite{Macey:1991,Isham:1993ji,Unruh:1976db,Hawking:1974sw}.
Suppose, at the classical level the evolution is governed by the constraint (more precisely,
a combination of spacetime diffeomorphisms' constraints) ${\cal H}  \approx 0$ which 
%is a function of the conjugated phase space variables and 
vanishes
in a weak sense.
In the quantum case one must define first a state
$\omega$: $\cal{A}\to \mathbb{C}$, a positive, linear
and normed functional on $\cal{A}$.
From the Gelfand-Naimark-Segal theorem it follows that there exists exactly one representation
$\pi_\omega$: $\cal{A}\to \text{End} \cal{H}_\omega$
of the algebra $\cal{A}$ on a Hilbert space $\cal{H}_\omega$, and the vector
$\xi_\omega \in \cal{H}_\omega$ such that:
(i) $\text{Lin}(\pi_\omega(\cal{A})\xi_\omega)=\cal{H}_\omega$,
(ii) $\omega(a) = (\pi_\omega(a)\xi_\omega,\xi_\omega)$ for every $a \in \cal{A}$ \cite{Murphy:1990}.
Then the evolution of physical states is governed by an appropriately chosen operator $\hat {\cal H}$.
% which is usually the generator of the coordinate-time translations.

To begin with, by $| \Psi \rangle \in \cal{H}_\omega$ we denote the wave functional which describes the state of the dynamical system.
Then in the generalized \schrod picture the quantum evolution equation can be written in some representation as (we consider pure states for simplicity):
\be\lb{e-qe}
\left[
%\hat{\cal{H}}_\text{tot} =
\hat{{\cal H}}  + F (\Psi)
\right]
 \Psi = 0,
\ee
where  
%$x$ and $p$ denote spacetime coordinates and conjugated momenta, respectively.
the first term in brackets is essentially the above-mentioned combination of constraints
${\cal H}$ quantized as in the conventional formalism.
Its explicit form is determined by a concrete physical setup
and thus will not be important for us
here -
we simply assume that it can be consistently defined.
This will make our following results largely model-independent from
spacetime-formulated theories.
The other term, $F(\Psi)$, is not present in the
conventional quantization procedure.
As to preserve the probabilistic interpretation of $\Psi$ (the physical states are actually not vectors but rays),
we assume that $F$
depends not on $\Psi$ alone but rather on its complex square.
Notice also that this term does not interfere with ${\cal H}$
as it describes the self-interaction of the wave functional and
thus is inherent only to the way we define the quantum wave equation.
Therefore, we write Eq. (\ref{e-qe}) as
\be\lb{e-qef}
\left[
\hat{\cal H} + F(\rho)
\right]  \Psi  = 0,
\ee
where $\rho \equiv |\Psi|^2$.
What is the explicit form of the operator $\hat F$?

%For clarity we will work in the position representation.
Suppose the system described by $\Psi$ consists of two separated distinguishable subsystems,
described by wave functionals $\Psi_1 $ and $\Psi_2 $,
respectively,
which obey the wave equations
$
%\be\lb{e-qefi}
%\hat{\cal{H}}_\text{tot}^{(i)}
\left[
\hat{{\cal H}}_i  + F(\rho_i)
\right]
 \Psi_i  = 0, \ \ i=1, 2,
$
where we denoted
$
\rho_i \equiv |\Psi_i|^2
$.
We assume that $F$ has the form
\be\lb{e-fro}
F(\rho) = - \beta^{-1} \ln{( \Omega \rho)},
\ee
where $\beta$ and $\Omega$ are arbitrary positive constants with the dimensionality of the inverse energy
and space volume, respectively.
This function
is a general solution of
the following algebraic equation
\be\lb{e-ff}
F(\rho_1 \rho_2)
=
F(\rho_1) + F(\rho_2),
\ee
if $\beta$ has the same value for both subsystems,
therefore, for two uncorrelated subsystems,
when the wave functional of a whole system becomes the product
$|\Psi \rangle = | \Psi_1 \rangle \otimes | \Psi_2 \rangle$,
the overall quantum wave equation turns to:
%\bw
$%\be\lb{e-post}
%\hat{\cal{H}}_\text{tot} (\Psi_1 (x^{(1)}) \Psi_2 (x^{(2)})) =
\left[
\hat{\cal{H}}_{12} 
+
F(\rho_1) + F(\rho_2)
\right]
(\Psi_1  \Psi_2 ) = 0$,
%\ee
%\ew
thus the non-linear part in this special case becomes a plain sum of the subsystems' ones.
Therefore, the non-linear term obeying Eq. (\ref{e-ff}) introduces on its own
no additional correlations between the uncorrelated subsystems for which $\beta$
takes the same value.
Later it will be shown that, for instance, in the weak-gravity limit
the relative difference between any two $\beta$'s is given by
$\Delta \beta / \beta \sim \Delta E/E_{\text{QG}} $,
where
$E_{\text{QG}}$ is certain very large energy scale
and
$\Delta E \ll E_{\text{QG}}$ is the difference of the subsystems' energies,
and therefore, in that limit Eq. (\ref{e-ff}) holds with a high degree of precision.
Thus, the logarithmic non-linearity on its own does not break the energy additivity and separability of non-interacting subsystems in non-relativistic quantum mechanics.
Therefore, from the viewpoint of the correspondence principle
the logarithmic term is safe to include into the quantum wave equation.

Gathering all together, from Eqs. (\ref{e-qef}) and (\ref{e-fro}) in
the position representation we obtain the following quantum wave equation:
\be\lb{e-xmain}
\left[
\hat{{\cal H}} 
-
\beta^{-1} \ln{(\Omega |\Psi(x)|^2)}
\right]
\Psi(x)
= 0,
\ee
which is the quantized version of the constraint
${\cal H}  \approx 0$ in generally covariant theories.
This equation can be formally written as
\be\lb{e-main}
\hat{\cal{H}}' |\Psi \rangle
\equiv
\left(
\hat{{\cal H}} 
+
\beta^{-1} \hat S_{ \Psi}
\right)
|\Psi \rangle = 0,
\ee
where $\hat S_\Psi$ is the Hermitian operator defined in the
following way \cite{ismk}:
let us consider the operator
$\hat S_{f}$ labeled by $f$,
$f$ is an arbitrary square-integrable
and
nowhere-vanishing
function, which satisfies the equation
$
\hat S_{f} |\Psi \rangle = -
 \ln{(\Omega |f|^2)} |\Psi \rangle
$.
It is easy to see that $\hat S_{f}$ is a quantum-mechanical operator in the conventional sense.
Then the operator $\hat S_\Psi$ is defined as
$\hat S_{f}$ evaluated on the  surface $f - \langle x |\Psi \rangle = 0$.
This implies the use of the projective Hilbert space but the latter is not a problem
because even in the conventional quantum formalism the space of physical states
is already a projective Hilbert space (rays), due to the normalization constraint.
Of course, to preserve interpretation of $\Psi$ as a wave function,
it must be normalizable and also the corresponding
probability density   must obey the conservation law
which is the case for the logarithmic nonlinearity \cite{BialynickiBirula:1976zp}.
Otherwise, the nonlinear quantum wave equation can be
interpreted only as an effective one, one example to be the Gross-Pitaevskii equation
(yet the term ``effective'' should be treated with care here because in quantum gravity, unlike
condensed matter, the background medium cannot be eliminated).

To summarize, the nonlinear term (\ref{e-fro})
has a number of physically meaningful distinctive properties which
make it a very probable candidate for a nonlinear correction to the quantum wave equation:
(i) it would not break the general covariance of a classical theory (by which we understand here a
coordinate-independent formulation of physics laws at the classical level),
%because it does not interfere with the spacetime diffeomorphisms' constraints (yet, of course, it takes them into account - through the dependence on the wave functional which, in turn, is a solution of the wave equation),
(ii) it does not violate the locality of product states
via EPR-type correlations, as discussed above \cite{Czachor:1997pw},
(iii) if wave function satisfies Eq. (\ref{e-xmain}) then this function times any constant
is also a solution provided that $\hat{\cal{H}}'$ is shifted by a constant,
(iv) symmetry properties of wave functions with respect to permutations of the
coordinates of identical particles are not affected by the nonlinear term,
%the \schrod equation for multicomponent wave functions can be easily written using
%the property Eq. (\ref{e-ff}),
(v) in the non-relativistic limit
%the properties of the total integrated force and torque about any center, as well as
the equations for the
probability density and current are not altered, energy is additive and
bounded from below,
free-particle waves are localized.
The non-linear corrections of other kind can also be included into Eq. (\ref{e-main})
but they would violate these properties, therefore, in what follows
we focus only on the non-linear term (\ref{e-fro}).
Below it will be shown that it has some additional properties which
make it very useful in quantum gravity theories.

\scn{Modular group and evolution time}{sec-phy}
The physical meaning of the non-linear term in Eq. (\ref{e-main}) becomes a bit more
clear when we go to the non-relativistic limit.
Its averaging in the position basis yields
%\bw
\be
\langle \beta^{-1} \hat S_\Psi \rangle \equiv -
\beta^{-1} \int
|\Psi|^2 \ln{(\Omega |\Psi|^2)}  d^3 x
% \nn\\&&\quad - \beta^{-1} \int  \rho (x) \ln{\rho (x)}\, d^3 x + \text{const}
 \equiv T_\Psi S_\Psi
%+ \text{const}
,
\ee
%\ew
where the functional $S_\Psi$ is the Shannon-type entropy
which can be formally assigned to a single quantum particle,
$T_\Psi \equiv (k_B \beta)^{-1}$, $k_B$ is the Boltzmann constant.
%in what follows we will work in units where $k_B =1$.
From the analysis of its discrete counterpart one can infer that $S_\Psi$ reaches a minimum
on the delta-like probability distribution
(which corresponds to a classically localized particle)
and maximizes on the uniform one \cite{ismk}.
This is what the information entropy usually does,
therefore,
$S_\Psi$ can be interpreted as a measure of the particle's ``smearing''
over space and corresponding quantum uncertainty.
Such quantum-mechanical entropy is a purely information one and
should not be confused with the von Neumann entropy -
the latter would vanish for pure states  and thus
neglect an inherent quantum-mechanical uncertainty of the outcome
of a measurement \cite{Jaynes:1957zz}.
Perhaps, the notion of entropy closest to $S_\Psi$ would be the one proposed
by Wehrl for the coherent states \cite{Wehrl:1979}.

Further, strictly speaking, Eq. (\ref{e-main}) contains no evolution time parameter.
It is the well-known property
of fully covariant theories that ${\cal H} $ contains the geometrical time-like coordinate
at most but the
dynamics cannot be formulated in terms of a single external time parameter.
This creates the conceptual difficulties because the full dynamics of the quantum
gravity theory can not be consistently defined in terms of other available notions of time,
such as the proper time, because they are state-dependent \cite{Connes:1994hv}.
They can serve as the evolution times only for dynamical theories on a
{\it fixed} spacetime geometry but not for the dynamical theory of geometry which the quantum gravity is supposed to be.

One way to define the evolution time without invoking assumptions for spacetime geometry
is to formulate it based on notions of a statistical and thermodynamical nature.
To our knowledge, the initial idea was proposed in
Ref. \cite{Rovelli:1993ys} - to define the evolution time as the vector flow of the Gibbs state
on the constraint surface.
Is there any way to derive the notion of the evolution time from the quantum wave equation itself?
We showed before that the nonlinear term from Eq. (\ref{e-main})
can be interpreted as a kind of entropy, at least in the non-relativistic limit.
Our next step will be to show that this term gives rise also
to the evolution time in quantum theory.

Luckily, the mathematical background for justifying this
has been already developed, both for the
conventional spacetime geometry \cite{Connes:1994hv} and
for the non-commutative one \cite{Heller:1997sj}.
Recalling the notations above, let ${\cal R}$ be the von Neumann algebra generated by the representation of ${\cal A}$, $\pi_\omega ({\cal A})$, on the Hilbert space.
Then
the Tomita-Takesaki theorem asserts that the mappings of
the von Neumann algebra on itself,
$\alpha_\tau$: ${\cal R} \to {\cal R}$ ($\tau \in \mathbb{R}$),
of the form
$
\alpha_\tau (b) = \Delta^{-i \tau}  b \Delta^{i \tau}, \
b \in {\cal R},
$
where $\Delta$ is a self-adjoint positive operator,
form a one-parameter group of automorphisms of ${\cal R}$, called the modular
group of the state $\omega$. 
This can be equivalently written as
\be\lb{e-delco}
\dot b \equiv \frac{d}{d \tau} \alpha_\tau (b)|_{\tau =0} = i [b, \ln \Delta],
\ee
thus, the ``modular'' time $\alpha_\tau$ can be already regarded as the time we are needed in
but it seems still state-dependent so a bit of final tuning must be made.
Some of automorphisms on ${\cal R}$ are inner-equivalent, the set of their equivalence classes forms a group of outer automorphisms $\text{Out}({\cal R})$.
Then the modular group $\alpha_\tau$ of a state $\omega$ projects down to a nontrivial group
$\tilde\alpha_\tau \in \text{Out}({\cal R})$.
The state-independent characterization of time is then proven by the cocycle Radon-Nykodym
theorem from which it follows that $\tilde\alpha_\tau$ does not depend on choice of $\omega$
in ${\cal R}$.

Further, from Eq. (\ref{e-main}) we obtain 
\be\lb{e-twocomm}
0 = [b, \hat{\cal{H}}']
=
[b, \hat{{\cal H}}]
+  \beta^{-1} [b,  \hat S_\Psi]
,
\ee
for any
$b \in {\cal R}$.
Then,
observing that the product
$\text{e}^{i \tau  {\cal{H}}'} b\, \text{e}^{-i \tau {\cal{H}}'}
$
is equivalent to
$
\text{e}^{i \tau  S_\Psi/\beta} b\, \text{e}^{-i  \tau S_\Psi/\beta }
$
on the constraints' surface,
we can assume that
\be\lb{e-time}
\Delta \propto
\exp{\hat S_\Psi}
,
\ee
from which, using Eq. (\ref{e-delco}), we obtain
the equation of motion in the generalized Heisenberg picture:
\be\lb{e-evo}
\frac{i}{\beta} \frac{d}{d \tau} b = [\hat{{\cal H}}, b],
\ee
where 
%$\hat{{\cal H}} = \hat{{\cal H}}(\hat x, \hat p)(\tau)$ and 
the
derivative is understood in a sense of the commutator from Eq. (\ref{e-delco})
with the generator given by Eq. (\ref{e-time}).
This essentially means that the dynamics described by the ``stationary''
operator $\hat{{\cal H}}'$
is equivalent to the evolution governed
by $\hat{{\cal H}}$ with respect to the evolution time $\beta\tau$.
In other words,  the logarithmic nonlinearity can be ``used up'' for creating
the time evolution of a generally covariant theory:
theory containing such nonlinearity but without
the observer-independent time evolution is
equivalent to the \textit{linear} theory
with the evolution time defined by the modular group.
As shown below, only in the low-energy limit this time parametrization becomes global.

%\newpage

\scn{Dispersion relations and observational tests}{sec-exp}
From Eq. (\ref{e-main}) one can see that the dispersion
relation for a particle in vacuum is being deformed by the non-linear term.
This is not surprising though as the quantum gravity is expected to give rise to
such corrections \cite{Ellis:1992eh,Garay:1998wk,AmelinoCamelia:1997gz}
because the gravitational medium contains quantum fluctuations which
respond differently to the propagation of particles of different energies -
the phenomenon somewhere analogous to propagation through electromagnetic plasmas \cite{Latorre:1994cv}.
%However, the most elegant way to see what is happening in our case is the following.
The full treatment of this problem is impossible without establishing
a concrete quantum gravity model.
Yet, some features are model-independent and can be clarified already at this stage.

First, despite the functional form of the non-linear term is universal for all dynamical systems
the constant $\beta$ is not a fundamental one hence depends on dynamical characteristics
of a system.
To find its physical meaning,
we go to the flat-spacetime limit and
consider a norm-preserving splitting of an arbitrary wave function into
$N$ non-overlapping parts of the same form as the initial wave function:
$
\Psi(x) \to \sum\limits_{i=1}^N \sqrt{p_i} \,\Psi(x - x_i)$ where
$
\sum\limits_{i=1}^N p_i = 1
$.
Then from the averaged Eq. (\ref{e-main}) one obtains that the change of energy during
such process amounts to
$
\delta E =
- \beta^{-1} \sum\limits_{i=1}^N p_i \ln{p_i}
$,
and thus $\beta$ is a measure of the binding or decoupling energy,
\be
\beta \propto 1/\delta E.
\ee
Second, from Eq. (\ref{e-evo}) one can immediately see that
%the single evolution time parameter is $\tilde\tau = \beta \tau$, therefore,
for any two dynamical
systems
the relation
\be
\beta_1 d \tau_1
%= d \tilde\tau
=  \beta_2 d \tau_2
,
\ee
must hold,
thus, if two systems have different $\beta$'s then their evolution time scales must differ as well.

Now, suppose that some two particles are products
of the reactions happened inside a compact region of space.
In the process these particles receive certain amounts of energy,
$E_1$ and $E_2$, respectively.
From the previous two equations we obtain that the ratio of their evolution time scales
is given by
\be\lb{e-b12}
\frac{d\tau_2}{d\tau_1}
=
\frac{\beta_1}{\beta_2} = \frac{E_2 - E_0}{E_1 - E_0}
=
1
-
\frac{E_2 - E_1}{E_0}
+
{\cal O} (E^2/E_0^2)
,
\ee
at least in the leading-order approximation.
Here $E_0$ is the energy of vacuum of a theory,
%here we assume it to be negative (as to avoid superluminal propagation afterwards), and also 
we imply that
$|E_0| = E_{\text{QG}}$ where $E_{\text{QG}} \lesssim 10^{19}$ GeV is
the effective quantum gravity energy scale.
The large value of the latter
explains why our (non-relativistic) notion of time is global
and energy-independent:
in the low-energy regime the
characteristic energies of any two particles become small comparing to the
quantum gravity energy scale, therefore, the difference between any two $\beta$'s
also gets vanishingly small,
$(\beta_2 - \beta_1)/\beta_i \sim (E_2 - E_1)/E_{\text{QG}}$,
and when the value of $\beta$ is essentially the
same for all dynamical systems then this constant can be absorbed into the time parameter.
Thus, the reason why the logarithmic nonlinearities are not observed in current
quantum-mechanical experiments
is not because of their smallness but because they act as the time derivatives
with individual scale factors which are essentially indistinguishable in the low-energy regime.

Another observation can be made if one recalls that
the correspondence principle implies that for elementary particles
their ``modular'' times must synchronize with their proper times in the weak-gravity
limit.
Then Eq. (\ref{e-b12}) immediately
reveals the presence of the Lorentz invariance violation (LIV):
it is not the conventional line element of spacetime
which is
invariant but the one multiplied by an energy-dependent function \cite{Asanov:1986aj}.
%As long as the energy itself is a nontrivial function of the coordinates and their incremental separations,
%such spacetime should be described in general case within the framework of the Finsler-Riemann geometry (suitably generalized for Lorentzian manifolds)
%In this geometry the metric structure is based on the line element
%$d s^2 = F^2(x^1, ..., x^D; d x^1, ..., d x^D)$
%where $F(x; y)$ is the function on the tangent bundle which is homogeneous of degree one
%in $y$ but does not necessarily obey the quadratic restriction \cite{Rnd:1959}.
However, in the low-energy limit this circumstance does not bring any phenomenological difficulties because
%then the energy-dependent factor
%gets evaluated on particles' world lines and becomes a function of the geometrical time and
for a conformally flat spacetime
(such as the Friedmann-Lemaitre-Robertson-Walker one) one can still define the representations
of the Lorentz group in the vierbein basis, therefore, the main notions of particle physics are
preserved \cite{Birrell:1982ix}.
In any case, LIV is an expected phenomenon in quantum gravity - the nontrivial
vacuum creates a preferred frame of reference.

Further, our particles fly off, travel across the space
and eventually get caught by a remote detector.
If they are of same kind, initially were emitted approximately simultaneously
with equal velocities
and their travel conditions were similar then
what kind of differences between them are our detectors supposed to catch?
Once again, the exact quantitative results are impossible without employing
the concrete model of quantum gravity yet some heuristic analysis 
can be done in the flat-spacetime approximation
(the cosmological corrections are not considered for now as to avoid certain confusions).

Making in Eq. (\ref{e-b12}) the transition from proper time to the
distant observer coordinate time $t$ (with cosmological effects 
%are taken into account
that would be the comoving time),
we obtain
\be\lb{e-vbra}
\frac{v_1}{v_2}
\sqrt{
\frac{c^2 - v_2^2}{c^2 - v_1^2}
     }
= \frac{E_2 - E_0}{E_1 - E_0}
,
\ee
where $v_i = d x/ d t_i$.
%In what follows we will be interested in their ratio $v_1/v_2$ as a function of $v_2$ and energies (which will be later used for deriving the dispersion relations).
This equation reveals the following subtlety: if for our future purposes
we assume that the particles are essentially relativistic 
or even ultrarelativistic and also that their velocities are nearly the same,
then the value of a square root in the equation
above crucially depends on whether the ratio $v_1/c$ tends to unity ``stronger'' than
$v_1/v_2$.

Thus, there exist  two limit regimes of analysis:
{\it linear} or standard relativistic - when the ratio
$v_1/v_2$ approaches one ``stronger'' than $v_1/c$ does, 
and {\it non-perturbative} or extreme ultrarelativistic - when it is other way around.
When analyzing concrete experimental data, 
one should look at these two ratios 
to decide which regime
s/he is next to.
Especially one should be careful when the cosmological effects 
are taken into account because 
%energy and velocity of the  
these ratios
may vary
as the particles propagate.

\sscn{Linear regime}{sec-regst}

In this case the square root in Eq. (\ref{e-vbra})
can be well approximated by one,
therefore,
under the above-mentioned assumptions
we  obtain:
\be\lb{e-v1v2}
\frac{v_1}{v_2} \approx \frac{E_2 - E_0}{E_1 - E_0}
,
\ee
thus, under equal conditions the particle with lower energy travels faster,
so we can write
$v \sim v^{(0)} / (1 - E/E_0)$ where 
%the value 
$ v^{(0)} = v_{E/E_0 \,=0}$. 
%tends to what would be the velocity of a particle in the conventional theory.
The difference in their arrival times is proportional to their energy difference:
\be\lb{e-times}
t_{2} - t_1
\approx
t_1
\frac{E_2 - E_1}{E_1 - E_0}
.
\ee
For instance, for the photons
we obtain
%\be\lb{e-phot}
$
\Delta t
\approx
\frac{L}{c E_{\text{QG}}} \Delta E
%\left[ 1 - \frac{E_1}{E_{\text{QG}}} + {\cal O} (E^2/E_{\text{QG}}^2) \right]
,
$
where $L$ is the distance from a distant observer to the actual place where
the reactions happened. 
This is 
what is being often observed about cosmic ray photons coming from the very distant Gamma-ray bursts
(GRB), the highly energetic explosions of massive stars in galaxies:
in the linear approximation
the difference in the arrival times of photons is proportional to their energy difference.
For instance, during the exceptionally luminous GRB 080916C, distant from us as far as 12.2 billions light years, the first photons with energies above
1 GeV started to arrive only after ten seconds after the trigger, e.g.,
the 13.2 GeV photon had
arrived after 16.54 s (the duration of the whole event itself was few tens seconds) \cite{FC:2009}.
The cosmological-scale remoteness of this and some other GRB's plays a crucial role here:
from the last formula it is clear that
$L/c$ should be very large so as to win over the huge number $E_{\text{QG}}/\Delta E$
%in the denominator
and produce an experimentally detectable effect.

The observational predictions can also be formulated on language
of the (deformed) dispersion relations for particles {\it in vacuo}.
From Eq. (\ref{e-v1v2}) we obtain
\be
\frac{\Delta v}{\Delta E}
\equiv \frac{v_2 - v_1}{E_2 - E_1}
= \frac{v_2}{E_0 - E_2}
,
\ee
therefore, if
$ E \ll |E_0| \sim E_{\text{QG}}$
then
$d v / d E \approx \Delta v / \Delta E \sim - \xi v / E_{\text{QG}}$,
where we assume $\xi \equiv - \Sign{E_0} = \pm 1$.
It means that 
the velocity
can be written as an exponent of energy
and thus
is a linear function of $E$
in the leading order:
\be\lb{e-disp1}
v/c \sim \exp{(- \xi E/E_{\text{QG}})}
=
1 - \xi \frac{E}{E_{\text{QG}}} + {\cal O} (E^2/E_{\text{QG}}^2) ,
\ee
and same result can be obtained if we look for a Taylor series 
solution of Eq. (\ref{e-v1v2}):
$
v_i/c = \sum\limits_{n=0}^{N} a_n (E_i/E_0)^n,
$
where $N \geq 1$ is an approximation order,
$a_n$ are energy-independent constants to be determined.
The boundary conditions are determined by physics -
for instance, for the case of photons they would be: 
$v_i = c$ when $E_i/E_0\to 0$.
 
With Eq. (\ref{e-disp1}) in hands we can immediately recover the results of Refs. \cite{AmelinoCamelia:1997gz,Kifune:1999ex,Protheroe:2000hp}.
Indeed, the resulting dispersion relation for the rotationally invariant
case, $\partial E/\partial p = v (E)$, integrates to
\bw
\be\lb{e-ddr}
c^2 (p-p_0)^2
\sim E_{\text{QG}}^2
\left(
1
%+ \frac{\mu c^2}{E_{\text{QG}} }
- \xi \text{e}^{\xi E/E_{\text{QG}}}
\right)^2
=
 E^2
\left[
1 + \xi \frac{E}{E_{\text{QG}}}
+ {\cal O} (E^2/E_{\text{QG}}^2)
\right]
,
\ee
\ew
where $p_0$ is an integration constant.
In the limit $p_0 \to 0$ the relation reduces to the dispersion relation
for massless
particles in the effective quantum gravity
theories based on $\kappa$-deformations of Poincar\'e symmetries
in which the time coordinate does not commute with spatial ones
\cite{Lukierski:1991pn,AmelinoCamelia:1996gp}.
On the other hand, with the appropriate choice of the integration constant the last equation
can be approximated by (in high-energy units):
$
E^2 \simeq p^2 + m^2 - (E /E_{\text{QG}}) p^2,
$
from which one can find out that the kinematics of particle-production processes
(such as the photopion production $p + \gamma \to p + \pi$, etc.) will be affected -
the photopion-production threshold energy gets increased by the deformation
%\bw\be
$
E > 
\frac{(2 m_p  + m_\pi) m_\pi}{4 \epsilon}
\left[
1
+
\frac{(2 m_p + m_\pi)^2 \, m_\pi^2}{64 \epsilon^3 E_{\text{QG}}}
\left(
1- \frac{m_p^2 + m_\pi^2}{(m_p + m_\pi)^2}
\right)
\right]
,
$
%\ee\ew
where $E$ and $\epsilon$ are the energies of a proton and cosmic microwave background photon, respectively, $m_p$ and $m_\pi$ are proton's and pion's masses.
Similarly, for the electron-positron pair production process
$\gamma + \gamma \to e^- + e^+$ one obtains the threshold
$
{\cal E} > \frac{m_e^2}{ \epsilon } +   \frac{m_e^6}{8 \epsilon^4 E_{\text{QG}} }
,
$
where ${\cal E}$ is the energy of a traveling photon, $m_e$ is the electron mass.

\sscn{Non-perturbative regime}{sec-regnst}

As was mentioned earlier, 
if the propagation speed of a particle is very close to
$c$ then one can not approximate
Eq. (\ref{e-vbra}) by Eq. (\ref{e-v1v2}), and
therefore, the expression 
(\ref{e-disp1}) can not be valid in general.

There exist at least three ways of how one can derive here 
the correct expression for velocity as a function of energy.
First way is to assume 
$v_2 = v_1 + \Delta v$, 
$E_2 = E_1 + \Delta E$
in Eq. (\ref{e-vbra}),
expand the latter to a linear order w.r.t. $\Delta v$,
replace therein $\Delta$'s by their infinitesimal
counterparts and integrate in the spirit which led us
to Eq. (\ref{e-disp1}).
Second way
%, the most rigorous,
is to look for an approximate solution of Eq. (\ref{e-vbra})
in a series form 
$ 
v_i/c = \sum\limits_{n=-2 N}^{2 N} a_n (\epsilon_i)^n,
$
where 
$\epsilon_i \equiv \sqrt{E_i/E_0} \ll 1$
($E_0 \equiv - \xi E_\text{QG}$, as before, $\xi = \pm 1$), 
%but in this subsection we assume it nonnegative for simplicity),
$N \geq 1$ is an approximation order,
$a_n$ are energy-independent constants to be determined.
The third method \cite{etk} might look not as rigorous as the previous two but
it is
fast and
elegant: one should just write
Eq. (\ref{e-vbra}) in the form
$
\sqrt{
\frac{(c/v_2)^{2}-1}
     {(c/v_1)^{2}-1}
     }
=
\frac{1-E_2/E_0}{1-E_1/E_0}
,
$
where its solution becomes obvious.

All these methods lead to the same result:
the desired $v(E)$ is a solution of the
algebraic equation
$
\sqrt{(c/v)^{2}-1}
=
\sqrt{\chi^2-1}
(1-E/E_0)
$,
namely
\be\lb{e-vnpt}
v/c
=
\left[
1+
(\chi^2 -1)
\left(
1-
\frac{E}{E_0}
\right)^2
\right]^{-1/2}
,
\ee
where 
$\chi 
%= c/ v|_{E/E_0 \to 0}
$ is the emerging parameter
which value can not be determined from Eq. (\ref{e-vbra}) alone.
By construction this parameter does not depend on energy  of
a particle but may vary for different kinds of particles. 
From this expression one can directly
compute the effective refractive index of the vacuum.
In the Cauchy form it can be written as \cite{cauchy}
\be
n^2 
\equiv 
(c/v_\gamma)^2
=
%\sqrt{
1 + \cmo_\gamma 
\left[1+ {\cal M}(\omega) (\omega/2 \pi c)^2
\right]
,
\ee
where
$
\cmo_\gamma = \chi_\gamma^2 -1 
%\ll 1
$
and
${\cal M} (\omega)
=
(2 \pi c/\omega_0)^2
\left(
1
+
2 
\xi
\omega_0 / \omega
\right)$
are, respectively, the constant of
refraction
and
dispersion coefficient
of the physical vacuum,
$v_\gamma$ is the velocity of a photon,
$\omega 
%\equiv E_\gamma /\hbar
$ is the angular frequency of the electromagnetic wave,
$\omega_0 = |E_0|/\hbar$ is the proper frequency of the vacuum.
%$\pm = -\Sign{E_0}$.
All this confirms once again that 
the physical vacuum is the medium with non-trivial
properties which affects photons and other particles propagating through it,
and the effects grow along with particles' energies. 

The final dispersion relation can be obtained by subsequent
integration, as for Eq. (\ref{e-ddr}):
\be\lb{e-pemu}
p -p_0
=
-
\frac{E_0}{2 c \sqrt{\cmo}}
\biggl[
\Upsilon
\left(
\sqrt{\mu}
(1-E/E_0)
\right)
-
\Upsilon
\left(
\sqrt{\mu}
\right)
\biggr]
,
\ee
where 
we denoted
$ \Upsilon (x) \equiv  x \sqrt{1+x^2} + \text{arcsinh}\, x$,
$\cmo \equiv \chi^2 -1$, and
$p_0$ is the integration constant representing the momentum
of the background, it is convenient to work in the comoving frame
of reference where $p_0 = 0$.
It can be convenient also to eliminate $\chi$
from Eqs. (\ref{e-vnpt}) and (\ref{e-pemu}) to obtain
the expression for momentum as a function of energy and velocity:
\be\lb{e-pev}
p
=
\frac{E_0 - E}{2 c \, \Gamma}
%\left(1- \frac{E}{E_0}\right)
\biggl[
\Upsilon
\left(
\Gamma/
(1-E/E_0)
\right)
-
\Upsilon
\left(
\Gamma
\right)
\biggr]
,
\ee 
where by $\Gamma \equiv \sqrt{(c/v)^2 -1}$ we denoted the inverse Lorentz factor.
In fact, this formula is the replacement of the relativistic dispersion relation
$p = E v/c^2$ for ultrarelativistic particles with high energies $E \not\ll |E_0|$.
Notice also that, as long as the function $ \Upsilon (x)$ can be expanded
into infinite series of powers of $E/E_0$, our dispersion relations
are conceptually different from the currently popular polynomial ones:
as the ratio $E /|E_0|$ approaches unity the higher-order terms can not be neglected anymore.

The main feature of the non-perturbative solution is that it
indicates the existence of the different classes or
sectors,
depending on the value of $\chi$.
However, unlike the classical relativity,  
%the propagation speed can approach $c$ for a finite value of the particle's energy.
%Thus, unlike the classical relativity theory, 
sectors of the
``subluminal''
($v \leq c$)
and ``luminal''
($v = c$)
particles are not totally disconnected: the propagation speed
of the subluminal
particles can reach $c$ at {\it finite} energy. Among other things, this may cause
the transluminal phenomena in vacuum discussed below.

First mode, called standard or analytic, can be seen when $\chi \not= 0$. 
In this case Eq. (\ref{e-vnpt}) allows expansion into the Taylor series
w.r.t. energy:
\be\lb{e-vnpt-s}
\frac{v^{(s)} }{c_\chi}=
%\left[
1
+
\frac{\chi^2-1}{\chi^2}
\frac{E}{E_0}
+ 
\frac{(\chi^2-1)(\chi^2 - \frac{3}{2})}{\chi^4} \frac{E^2}{E_0^2}
+
{\cal O} (E^3/E_{0}^3)
%\right]
,
\ee
where $c_\chi \equiv c/\chi$ is the ``renormalized'' speed of light.
%in the physical vacuum.
This
shows that for photons in this mode the inverse of $\chi$ can be interpreted
as the ``luminal'' Mach number and thus $\chi$ is related to the (effective)
refractive index
of the physical vacuum.
%: $n \equiv c / c_\chi = \chi$.
To prevent their motion from becoming superluminal in this mode, 
$\chi^2$ must be larger than $1$
- but not  much larger, most probably no more than ten percent,
as to retain the formal value of $c_\chi$ close to $c$.
This makes the dimensionless series coefficients in Eq. (\ref{e-vnpt-s}) small
- in addition to the smallness of the ratio $E/E_0$ itself.

Another mode, called anomalous, is given
by the non-analytic branch of the solution (\ref{e-vnpt})
at $\chi = 0$:
\be\lb{e-vnpt-t}
v^{(t)} =
\frac{c/\sqrt{2}}{\sqrt{E/E_0}}
\left[
1
+
\frac{1}{4}
\frac{E}{E_0}
+
{\cal O} (E^2/E_{0}^2)
\right]
,
\ee
of course, this expression can be valid when
$v \grsim c$ where energy of a particle can not be vanishing.
For ultrarelativistic particles there is no obvious boundary 
condition to rule such modes out,
therefore,
%we conjecture that 
this mode is the (particular example of the)
``superluminal'' one:
it is an essentially LIV phenomenon which
%can be regarded as 
describes 
a particle 
which can propagate
in the (nontrivial) vacuum with the velocity larger than $c$. 

Generally speaking, the superluminal modes exist 
for  $\chi^2 < 1$, as one can directly see from Eq. (\ref{e-vnpt}).
Unlike the tachyons in the classical relativity theory, their energies are real-valued
and stay finite when $v$ approaches $c$.
If we extend $\chi$ on to the complex plane 
(but keeping its square real-valued)
then
the superluminal particles can be further classified depending on whether 
$\chi^2$ is a strictly positive number or not.
If it is then the minimal allowed energy of such particles
is zero, otherwise, i.e., when $\chi$ is imaginary or zero,
the energy must be greater than some threshold value:
$
%E^{(t)} > 
E_\text{min} \equiv E_0 (1 - 1/\sqrt{1-\chi^2})$.
Thus, they are not expandable in series in the vicinity $E=0$.
The mode (\ref{e-vnpt-t}) 
is a special case of the second subclass and is a kind of the ``interface''
mode between the subclasses: its $E_\text{min}$ is zero (similarly to the first subclass) 
but $v (E)$ is not analytic in that point (similarly to the second subclass). 

The common feature of the superluminal particles is 
for the ``sub-Planckian'' 
%range of 
energies ($|E/E_0| < 1$) their propagation 
speed 
decreases
as energy increases - as opposite to the subluminal mode (\ref{e-vnpt-s}) -
until it reaches $c$.
%The fact that Eq. (\ref{e-vnpt-t}) looks so unusual and differs so drastically from Eq. (\ref{e-vnpt-s}) 
This can be explained by 
when a particle propagates 
faster than the speed of light in the physical vacuum
%above the speed-of-light barrier 
its interaction with the latter 
leads to the ``luminal boom'' 
%(especially, when the velocity is nearly equal to $c$)
and
appearance of a conical front of the shock waves
%are necessary for 
carrying away large amount of energy. 
In the classical relativity this energy is actually infinite and thus the barrier crossing would be forbidden for known particles.
This Cherenkov-type radiation 
%in general can be %not only gravitational but 
is mostly electromagnetic but it can lead to
creation of other known particles - often with very high energies.
%where $\chi = \chi (v, E)$ is determined from Eq. (\ref{e-vnpt}).
An interesting question is whether the electromagnetic component
of this radiation
exhibit the anomalous Doppler effect - similar to the one
for the superluminal (non-point) sources in vacuum 
which has been predicted even at the level of the
classical relativity theory \cite{Bolotovskii:1972ve}. 
Of course, the vacuum Cherenkov radiation is an essentially  LIV
phenomenon \cite{Beall:1970rw,Coleman:1997xq,jlm2003,Lehnert:2004hq,Castorina:2004hv,Kaufhold:2005vj,Altschul:2006zz}: the relativistic superluminal point particles (tachyons)
can not emit it due to a trivial structure of the vacuum
there \cite{Mignani:1973pq}.

From the observational point of view, 
so far 
there exist not so much
arguments 
directly
supporting the existence of luminal booms in the (nontrivial)
vacuum in some GRBs and AGNs - 
probably because one can often find more than one way of explaining the majority of astrophysical phenomena, especially if a phenomenon is complex and/or the observational data lack of a necessary accuracy.
Nevertheless, some arguments do exist:
for instance, the softening of a GRB afterglow 
%\cite{maz82} 
is similar
to the frequency evolution of a sonic boom: 
at the surface of the shock cone the frequency is very high
but rapidly decreases inside,
and  there is even some quantitative similarity \cite{Thulasidas:2007hv}.
Also 
there may appear the phenomenon of
mimicking the double-lobed radio sources,
such as DRAGNs, by such shock waves -
by analogy with any two sound waves from a supersonic jet initially
emitted at different times (and thus from different locations)  
but reached an observer simultaneously thus creating 
an illusion of the doubling of the sound source.
In this connection, another interesting 
question is what would be the ``luminal'' analogues of other trans- and supersonic phenomena, such as the Prandtl-Glauert singularity, N- and U-wave, {\it etc}.

To summarize, all this 
%illustrates the fact that our semi-heuristic 
reasoning which led us from Eq. (\ref{e-b12})
to Eqs. (\ref{e-disp1})
%, (\ref{e-vnpt}) 
and (\ref{e-vnpt})
indicates
that 
in our theory
the dispersion relation 
%of a particle {\it in vacuo}
for all scales of energy and momentum
may actually vary
depending on a physical situation,
and therefore, 
%in our theory
the complete physical picture
is still on its way -
main reason of which was explained in the paragraph
preceding Eq. (\ref{e-vbra}).
In this connection, it would be interesting to find
the way of observational testing of directly
Eq. (\ref{e-b12}), or Eq. (\ref{e-vbra}) but taking into account
cosmological effects, as they are the primary predictions of the theory.

\scn{Conclusion}{sec-cnc}

The nonlinear extensions of quantum mechanics constantly 
attract attention for a number of reasons.
At first, one can not exclude the possibility that the conventional quantum mechanics is
just an approximate (``linearized'') limit of some theory which has much
wider range of applicability.
Thus, there is a strong hope that they can help solving the long-standing issues
of the conventional quantum theory and make a progress towards formulating
a satisfactory theory of quantum gravity.
On the other hand, up to now there is no mathematical proof or experimental evidence
whatsoever
which would rule out the nonlinear corrections
with a satisfactory degree of accuracy.
Moreover, 
there exist the physical situations in which
the nonlinearities might play the dominant role and thus must be accounted for.
For instance, there is a significant amount of experimental and theoretical
evidence that the physical vacuum possesses a non-trivial structure
which is an essentially non-perturbative and very likely a non-linear phenomenon.
In particular, this structure can cause the universal deformation
of dispersion relations and hence the quantum wave equations 
describing the elementary particles 
propagating through the vacuum.
While the microscopical theory of such nontrivial vacuum (known as the quantum gravity)
is still pending, some heuristic approaches can be very helpful, both
in terms of better understanding of an underlying theory and deducing the possible
phenomenological implications.

Nowadays, there exist two popular ways of formulating the extensions of the quantum mechanics while preserving the notion of a point particle as a fundamental one.
First way is to modify the commutation relations which leads to the non-commutative
quantum mechanics, q-deformed Heisenberg algebras, \textit{etc}.
Second is to directly deform the equations which determine the wave function itself.
The two approaches are not mutually exclusive although direct correspondences
are difficult to establish.
In this paper we
follow the second approach:
%within the framework of a generic generally covariant theory 
we introduce the logarithmic correction to the quantum wave equation. 
We advocate this kind of non-linearity for being minimal in a sense it introduces
the new physics yet, contrary to other candidates,
it preserves some important properties of the conventional quantum mechanics,
such as the separability, energy additivity and Planck relation.
We demonstrated the emergence of the evolution time from the group of automorphisms of the von Neumann algebra governed by this non-linear correction. It turns out that such time parameterization is essentially energy-dependent and becomes global only asymptotically - when the energies get very small comparing to the effective quantum gravity scale.
Similar thing happens to the Lorentz invariance: in the resulting theory it becomes an asymptotic low-energy phenomenon.

We also showed how the logarithmic non-linearity deforms the vacuum wave dispersion relations and explains certain features of the astrophysical data coming from recent  observations of high-energy cosmic rays.
In general, our estimates imply that
due to
the effects of the nontrivial physical vacuum 
%of quantum gravity
the mean free path of a subluminal high-energy particle, its lifetime in a high-energy state and, therefore, travel distance from the source can be significantly larger than one would expect
from the conventional theory.
In fact, using arguments of such kind
one can show that 
%Eq. (\ref{e-b12}) or, equivalently, 
the deformed dispersion relations above
are capable of explaining results of few other classes of experiments:
observations of cosmic rays above the expected GZK limit,
studies of the longitudinal development of the air showers produced by ultra-high-energy hadronic particles,
ATIC observations of the high-energy electrons from an unseen source
%(in the deformed theory %their lifetime in a high-energy state is significantly larger, thus
\cite{Kifune:1999ex,Protheroe:2000hp,AmelinoCamelia:2001qf,:2008zzr,Stecker:2009hj}.
The trans- and superluminal  phenomena 
in the nontrivial LIV vacuum are briefly discussed as well.

{\it Note added in proof}. 
After the initial version of this paper has
been e-printed we found that in the very recent experimental 
article, 
``A limit on the variation of the speed of light arising from quantum
gravity effects'' by 
the Fermi LAT and GBM Collaborations,
%A. Abdo {\it et al.},
%from the Fermi GBM/LAT Collaborations, 
(Nature 462, 331-334 [ArXiv:0908.1832]), the
velocity
dispersion (\ref{e-disp1}) has been ruled out.
For our theory it is not a problem though because 
in that particular case
%physical situation 
%they were gathering data from 
the linear approximation 
described in Sec. \ref{sec-regst} is hardly valid:
for the extremely 
ultrarelativistic particles, one should consider instead the 
exact dispersion relations from Sec. \ref{sec-regnst},
eventually leading
to Eqs. (\ref{e-vnpt}) and (\ref{e-vnpt-s}).
These are not ruled out by current 
observations - the Fermi's data can only put further bounds for 
the constant of refraction of the physical vacuum.

\begin{acknowledgments}
%\footnotesize

This article is based on a talk given at the Gamow'105 Memorial conference (Odessa, 2009).
I acknowledge the fruitful discussions with A. Avdeenkov, M. Kastner,  F. Petruccione, M. Porrmann, F. Scholtz, I. Snyman, E. Tkalya and
S. Vaidya, as well as the useful correspondence
from M. Czachor, N. Mavromatos, V. Mitsou and F. Stecker.
I am grateful also to C. Chryssomalakos, H. Quevedo and D. Sudarsky
from the Universidad Nacional Aut\'onoma de M\'exico
who introduced
me into the field of effective quantum gravity some time ago.
My research was partially supported by grant 68523 from the National Research Foundation of South Africa.
Last, but not least, I thank the
National Institute of Theoretical Physics (NITheP)
and
Stellenbosch Institute of Advanced Studies (STIAS)
and their Directors, F. Scholtz and H. Geyer, for hospitality.\\
%NITheP activities are being supported by a grant from the National Research Foundation of South Africa.

\end{acknowledgments}

\def\AnP{Ann. Phys.}
\def\APP{Acta Phys. Polon.}
\def\CJP{Czech. J. Phys.}
\def\CMPh{Commun. Math. Phys.}
\def\CQG {Class. Quantum Grav.}
\def\EPL  {Europhys. Lett.}
\def\IJMP  {Int. J. Mod. Phys.}
\def\JMP{J. Math. Phys.}
\def\JPh{J. Phys.}
\def\FP{Fortschr. Phys.}
\def\GRG {Gen. Relativ. Gravit.}
\def\GC {Gravit. Cosmol.}
\def\LMPh {Lett. Math. Phys.}
\def\MPL  {Mod. Phys. Lett.}
\def\Nat {Nature}
\def\NCim {Nuovo Cimento}
\def\NPh  {Nucl. Phys.}
\def\PhE  {Phys.Essays}
\def\PhL  {Phys. Lett.}
\def\PhR  {Phys. Rev.}
\def\PhRL {Phys. Rev. Lett.}
\def\PhRp {Phys. Rept.}
\def\RMP  {Rev. Mod. Phys.}
\def\TMF {Teor. Mat. Fiz.}
\def\prp {report}
\def\Prp {Report}

\def\jn#1#2#3#4#5{{#1}{#2} {\bf #3}, {#4} {(#5)}} %PRD
%\def\jn#1#2#3#4#5{{#1}{#2} {#3} {(#5)} {#4}}   %PLB style
% #1 tittle  #2 ser  #3 vol  #4 page  #5 year

\def\boo#1#2#3#4#5{{\it #1} ({#2}, {#3}, {#4}){#5}}
%\def\boo#1#2#3#4#5{ #1 ({#2}, {#3}, {#4}){#5}}  %PLB style
% #1 tittle  #2 publisher  #3 place  #4 year  #5 page/, p.789/

%\def\jn#1#2#3#4#5{{#1}{#2} {\bf #3}, {#4} {(#5)}}
% #1 tittle  #2 ser  #3 vol  #4 page  #5 year
%\def\boo#1#2#3#4#5{{\it #1} ({#2}, {#3}, {#4}){#5}}
% #1 tittle  #2 publisher  #3 place  #4 year  #5 page/, p.789/

%\newpage


\begin{thebibliography}{99}
\bibitem{cauchy}
Here we use the original definition of the Cauchy's formula,
in other versions
the square of
the refractive index is often omitted,
due to the smallness of the constant of
refraction, also the latter is rescaled by factor two.
\end{thebibliography}
\end{document}